\documentclass[11pt]{article}
\usepackage[latin1]{inputenc}
\usepackage[english]{babel}
\usepackage[namelimits]{amsmath}
\usepackage{amssymb}
\usepackage{amsmath}
\usepackage{amsthm}

\begin{document}
\title{Entropy, energy and temperature-length inequality for Friedmann universes}
\author{
Stefano Viaggiu,\\
Dipartimento di Matematica,
Universit\`a di Roma ``Tor Vergata'',\\
Via della Ricerca Scientifica, 1, I-00133 Roma, Italy.\\
E-mail: {\tt viaggiu@axp.mat.uniroma2.it}}
\date{\today}\maketitle
\begin{abstract}
In this paper we continue the study of the physical consequences of our modified black hole entropy formula in expanding
spacetimes. In particular,
we apply the new formula to apparent horizons of Friedmann expanding universes with zero, negative and positive spatial 
curvature. As a first result, we found 
that, apart from the static Einstein solution, the only Friedmann spacetimes with constant (zero) 
internal energy are the ones with zero spatial curvature. This happens because, 
in the computation of the internal energy $U$, the contribution due to the non-vanishing Hubble flow must been added to the 
usual Misner-Sharp energy giving, for zero curvature spacetimes, a zero value for $U$. This fact does not hold when curvature is present.
After analyzing the free energy $F$, we obtain the correct result 
that $F$ is stationary only for physical systems in isothermal equilibrium, i.e. a de Sitter expanding universe. 
This result permits us to trace back a physically reasonable hypothesis
concerning the origin of the early and late times de Sitter phase of our universe. 
Finally, we deduce an interesting temperature-length inequality similar to the time-energy uncertainty 
of ordinary quantum mechanics but with temperature instead of time coordinate. Remarkably, this relation is independent on 
the gravitational constant
$G$ and can thus be explored also in non gravitational contexts.
\end{abstract}
PACS number(s): 04.70.Bw, 04.70.-s, 04.70.Dy, 04.20.Cv

\section{Introduction}
A great amount of experimental observations during the past decade (see for example\cite{1,2,3}) 
are in agreement with the hypothesis of a present day accelerating universe. 
The standard 
$\Lambda$CDM cosmological model is described in terms of a spatially flat Friedmann metric endowed with a 
cosmological constant $\Lambda$ representing about $70\%$ of the present universe matter-energy content. 
Despite the enormous success of this model in explaining the cosmological data, some issues have to be furthert investigated:
origin of the dark energy, origin of the inflation, anisotropies of CMBR to cite the most studied ones. Another issue 
is due to the origin of the entropy of the universe. In the usual view of our universe, the Friedmann equations and the first law of
thermodynamics applied to comoving spheres imply an adiabatically expanding universe evolving with constant entropy $S$. As pointed in
\cite{5}, our universe does contain a large amount of entropy, and it is unclear how such a very large entropy was created at the big
bang or during the primordial inflation. In particular, the actual entropy $S_{obs}$ of the observable universe seems to be  
of the order $S/k_B\sim 3.1\times 10^{104}$, where $k_B$ is the Boltzmann constant (see \cite{6} and references therein). 
The supermassive black holes \cite{6} represent the largest contributor to the entropy of our universe. 
However, the entropy is a measure of the degrees of freedom of a system and in the calculation of $S_{obs}$  the contribution
due to the non static nature of Friedmann universes should be included. Unfortunately, 
the way to calculate the degrees of freedom of a dynamic universe is not yet
available.
Irrespective to this estimated value, it seems rather unlikely that such a huge amount of entropy was created soon after the big bang where
the proper dimensions of the observable universe was of the order of the Planck length $L_P$.

A very useful and intriguing way to understand the amount of the entropy of our observable universe is due to the holographic principle
(see \cite{7,8}). According to this principle, the maximum entropy of a given spherical space region is provided by the biggest black hole sitting
inside the region. In an expanding dynamical spacetime (see \cite{8a} and references therein) the identification of a black hole is not a simple task. 
In fact, the event horizon becomes a teleological concept non achievable for finite human observers. Fortunately, in references
\cite{10,11} it has been shown that the apparent horizon is a more suitable and manageable tool to identify black holes in an expanding universe.
Obviously, the identification between apparent horizons and black holes in a dynamical context is matter of debate and possible
drawbacks can emerge (see for example \cite{r1,r2,r3} and references therein). First of all, as well known (see for example
\cite{r1}), apparent horizons depend on the foliation of the spacetime. However, we are mainly interested to the apparent horizons of 
Friedmann spacetimes. In this regard, a given comoving observer is equipped with its apparent horizon of a given proper spherical radius
$L_h$, while a different comoving observer sitting in another galaxy at the same cosmic time $t$
will be equipped with a different apparent horizon (a spherical region
centered in its galaxy) with proper radius of magnitude $L_h^{\prime}$ but with $L_h=L_h^{\prime}$. Hence, 
any comoving experimenter in the universe agrees with the fact that
the universe can be considered as a thermodynamic object at its apparent horizon at the proper areal radius $L_h(t)$.  
Another possible drawback is due to the fact that dynamical horizons can be also timelike or null (see for example \cite{r2}), it 
substantially depending on the matter equation of state filling the universe. The same happens for apparent horizons in Friedmann spacetimes
\cite{4}. Finally (see \cite{r3} and references therein), some ambiguity can arise
for the definition of the surface gravity and related Hawking temperature for dynamical black holes. However, we stress again that we are interested to the study
of the thermodynamics of a Friedmann universe at its apparent horizon, without an embedded black hole.

In this regard, any Friedmann universe with positive energy-density has an apparent horizon. This apparent horizon can be equipped with a 
well-defined
surface gravity expressed in terms of the Kodama vector (see for example \cite{9a}).
As a consequence of these reasonings,
we may apply the holographic principle by supposing that the whole entropy of the universe at the apparent horizon with areal radius $L_h$
is given by the one of a black hole of radius $L_h$ and proper area $A_h$, i.e the celebrated Bekenstein-Hawking entropy
\cite{12,13} given by $S_{BH}=\frac{k_B A_h}{4 L_P^2}$. This formula gives a value for $S_{obs}$ given by $S_{obs}/k_B\simeq 10^{122}$ \cite{7,8} 
that is greater than the one estimated in \cite{6} without including the gravitational degrees of freedom due to the expansion of the universe that are 
not directly detectable from the visible and dark matter content of our universe. In practice, in order to calculate the contribution to $S_{obs}$
given by the dynamical degrees of freedom, a
Friedmann universe can be seen 'temporarily' (see \cite{nat}) as a black hole at its apparent horizon. 
More recently (see \cite{9,9a}), the holographic principle has been applied to the apparent horizon in order to 'read' the Friedmann equations
as representing exactly the first law of thermodynamics. Moreover, in \cite{r4} has been shown that the suitable notion of temperature is the one present in reference \cite{9a} and that the apparent horizon is the suitable radius to study the thermodynamics of the universe.
It is thus of great importance to study the validity of this identification 
in an expanding universe and its physical consequences.
 
We have recently argued \cite{8a,14} that after using the suitable theorems for the formation of trapped surfaces 
caused by spherical mass concentration in Friedmann universes (for istance see \cite{15,16,17,18}), the usual formula  
$S_{BH}=\frac{k_B A}{4 L_P^2}$ must be modified. Such a modification includes the Hubble flow contribution to the entropy of the universe. In this paper 
we continue the investigations present in \cite{8a,14} by analyzing the thermodynamcal consequences of our generalized entropy formula. 
In particular, we explore our formula for the internal energy and the free Helmholtz energy. This allows us to investigate the expansion history
of our visible universe in terms of the (on average) temperatue of the universe $T_u$ and the holographic temperature $T_h$ at the
apparent horizon. This way, we can depict the existence of a pure de Sitter phase of the universe in terms of the difference $T_u-T_h$.

In section 2 we present a derivation of our generilized entropy formula for Friedmann spacetimes. 
In section 3 we write down the first law of thermodynamics.
In section 4 we give the explicit expression of the internal energy $U$ (in terms of a time integration) and the effective pressure acting at the apparent horizon. In section 5 
we study the free Helmholtz energy $F$. 
In section 6 some reasonings concerning a  
 physically motivated relation between the holographic temperature $T_h$ at the apparent horizon 
 $L_h$ and the 'thermodynamical' termperature of the universe $T_u$ inside $L_h$ are given.
Finally, section 7 is devoted to some final remarks and discussions
while in
the appendix a we analyze our entropy formula for the McVittie solution and in appendix b
we study the flat case with a discussion concerning the partition function.

\section{Entropy of Friedmann universes}

In this paper we adopt the holographic philosophy to calculate the entropy of Friedmann universes. The line element is 
\begin{equation}
ds^2=-c^2 dt^2+a^2(t)\left[\frac{{dr}^2}{1-kr^2}+r^2{d\Omega}^2\right],
\label{1}
\end{equation}
where as usual $k=-1,0,+1$. In the flat case, we assume $a(t)$ adimensional, while in the other cases $a(t)$ is a length and both 
$k$ and $r$ are dimensionless. In \cite{14} we have analyzed the flat case and the hyperbolic and closed universes are briefly analyzed in
\cite{8a}. As it is well known, all Friedmann universes satisfying weak energy condition have an apparent horizon at the proper areal radius
$L_h$ given by
\begin{equation} 
L_h=\frac{c}{\sqrt{H^2+\frac{k c^2}{a(t)^2}}}.
\label{2}
\end{equation}  
Only for the Milne solution without any matter content and $k=-1$ we have $L_h=\infty$.

The core of the holographic principle \cite{7,8} is that the maximum entropy $S$ available in a region of size $R$ and area $A$
is the one of a black hole of the same dimensions, i.e. the largest black hole fitting inside.
Since the celebrated Bekenstein-Hawking (BH) formula \cite{12,13}, it is customary to pose $S\leq S_{max}=S_{BH}=\frac{k_B A}{4 L_P^2}$. 
This expression can be 
derived thanks to the Bekenstein bound or the entropy bound (see \cite{8a} and references therein): 
there exists an universal bound for the
entropy $S$ of a spherical object of radius $R$ and mass-energy $E$ given by 
$S\leq S_{max}=\frac{2\pi k_B RE}{\hbar c}$. By setting $E=c^4 R/(2G)$, the BH formula arises. Hence, in the static asymptotically flat case the
bound is satured with $E=M c^2$, where $M$ is the ADM (total) mass of the black hole.
In the literature (see for example \cite{5,9,9a}) the usual expression $S\sim A/4$ is also used  in a cosmological non-static context.
This line of research is motivated by the request to recast the Friedmann equations as the first law of thermodynamics.
However, it should be stressed that the 
expression $E=c^4 R/(2G)$ is certainly suitable for static black holes in asymptotically flat spacetimes, but it is rather questionable
in a cosmological context. Physical intuition suggests that in an expanding universe as the one we live, it is more difficult 
to build a black hole since the expansion is in opposition to the mass-concentration. In fact, theorems quoted in
\cite{15,16,17,18} confirm the physical intuition. These theorems are the starting point of our study. Thanks to the holographic principle, 
it is natural to suppose (see also \cite{8a}) that
the entropy bound can be exactly saturated at the apparent horizon (see also \cite{9,9a})
given by (\ref{2}). In practice, a Friedmann universe is temporarily a black hole \cite{nat} at its apparent horizon.
We must take a Friedmann solution 
and wonder what is the mass concentration $\delta M$ that leads to the formation of a trapped surface. 
The theorems present in \cite{15,16,17,18} are exactly what we need.\\
We denote with $\Sigma$ the spherical three-dimensional hypersurface obtained at $t=const$.
The initial data set $g_{ab}, K_{ab}, \rho$ are given on $\Sigma$,  
$g_{ab}$ is the metric, $K_{ab}$ the extrinsic curvature and $\rho$ the energy-density of the spherically perturbed 
Friedmann flat spacetime on the three-dimensional space $\Sigma$ by means of the mass excess $\delta M$.
We  also suppose that the matter current $J_a$ flowing through $\Sigma$ is vanishing.
Suppose that $K_{a}^{a}=const$, i.e. the rate of change of the volume of $\Sigma$ is not perturbed: if for a sphere $S$
with proper mass excess $\delta M$ we have
\begin{equation}
\frac{G}{c^2}\delta M < \frac{L}{2}+\frac{AH}{4\pi c}-k\frac{3V}{8\pi a(t)^2},
\label{3}
\end{equation} 
then $S$ is not trapped. After applying the entropy bound but with $E=\delta M c^2$ and $\delta M$ given by the second member of (\ref{3})
we obtain for the entropy $S_h$ at $L_h$ (see \cite{8a})
\begin{equation}
S_{h}=\frac{k_B A_h}{4 L_P^2}+\frac{3k_B}{2c L_P^2}V_h H-
\frac{3k k_B}{4L_P^2}\frac{L_h V_h}{a(t)^2},
\label{4}
\end{equation}
where as usual (see \cite{19,20}) we take, also for $k=-1,+1$, $V_h=4\pi L_h^3/3$. 
The application of (\ref{4}) to flat and hyperbolic expanding spacetimes gives no problem
(see also appendix a for an application to the McVittie solution). 

Some issue can arise when one applies the formula (\ref{4}) for closed universes with $k=+1$ that are contracting
$H<0$. In particular, when $L_h=a(t)$ we have $S_{BH}=0$ and the entropy becomes negative when $H<0$. The same phenomenon appears in
\cite{21}, in the context of the Cardy-Verlinde formula. This fact,
as pointed in \cite{21}, could indicate a problem  for the entropy bound  for closed 
universes near the turning point where $H_{min}=0$ and for $H<0$. Note also that for $k=+1$ and near the turning point at $L_h=a(t)$, 
the leading term
of (\ref{4}) is exactly the one given by the Cardy-Verlinde formula, provided that the bound (see \cite{21})
$S\leq S_{max}=\frac{2\pi k_B RE}{3\hbar c}$ (in the ordinary spacetime with dimensions $D=3+1$) is used.
In our context, as noticed in \cite{18}, trapped surfaces naturally arise in a closed universe at the so called south pole and
everywhere in the contracting phase, independently on the mass excess $\delta M$. It is therefore evident that in these cases  
the inequality (\ref{3}) can be satisfied also for $\delta M<0$ and this makes our procedure to obtain the expression (\ref{4}) questionable.
To avoid these shortcomings, we apply our new formula (\ref{4}) to expanding universes with $k=-1,0$ (independently on their matter content)
and to closed monotonic ones, for example equipped with a positive cosmological constant, or
to a cyclic universe in the expanding phase but with $L_h<<a(t)$. 

Note that for the derivation of (\ref{4}) in expanding universes it is assumed $a_{,t}/a(t)\neq 0$, i.e. it applies to dynamical universes.
Hence, to obtain the static limit we must impose from the onset $H=0$ and as a consequence the entropy of a static de Sitter universe looks like
the usual expression $S\sim A/4$.

We must specify the expression of the temperature $T_{D_h}$ at the apparent horizon. The usual expression is the one proportional to the surface gravity
parameter of a Kodama vector field (see \cite{8a}):
\begin{equation}
T_{D_h}=\frac{\hbar c}{2\pi k_B L_h}\left |1-\frac{L_{h,t}}{2H L_h}\right |.
\label{5}
\end{equation}
However, according to the usual thermodynamic, we have ${\left(\frac{\partial S}{\partial U}\right)}_{V_h}=1/T$ and thus the 'thermodynamic'
version of (\ref{5}) must be formally calculated at fixed $L_h$. As a consequence, the temperature appropriate in our context is
\begin{equation} 
T_{h}=\frac{\hbar c}{2\pi k_B L_h}.
\label{6}
\end{equation}
We denote the temperature $T_h$ as the holographic temperature at $L_h$.
  
\section{First law of thermodynamics}
We are in the position to derive the first law of thermodynamics. As customary, we can obtain the first law from the finite expression 
(\ref{4}) by calculating the expression $T_h dS_h$ where (see \cite{8a})
\begin{equation}
T_{h}dS_h=dU_h+P_h dV_h
\label{7}
\end{equation}
where $T_h$ is the temperature of the apparent horizon,
$U_h$ denotes the internal energy and $P_h$ the effective holographic pressure acting on $L_h$. Since \cite{9,9a,19}, 
the apparent horizon is dynamical, the expression for $P_h$ is not generally expected to be the one $p$ effectively present in the energy momentum tensor, 
i.e. $|P_h|\neq |p|$. 
In the usual picture \cite{9a}, the Friedmann equations are equivalent to (\ref{7}) by setting 
$S_h\sim A_h/4$, $U_h=E_{ms}$ where $E_{ms}$ is the Misner-Sharp energy
and $P_h=(p-\rho c^2)/2$. Note that in that case $P_h=0$ for stiff matter 
(although the apparent horizon is dynamical also in this case) and the work
term of the visible universe is negative for expanding universes filled with ordinary matter satisfying the dominant energy condition.
In the following we show that this picture is modified thanks to the new
physically motivated formula (\ref{4}).

To start with, note that irrespective of the fact that in the Friedmann case all the above quantities depend only on the cosmic comoving time $t$, 
we use the practical rule to identify the terms proportional to $dV_h$ as $P_h$ and the remaining terms to $dU_h$. 
It should also be noticed that after the transformation 
$T_h\rightarrow \beta T_h$, $U_h\rightarrow \beta U_h$ and $P_h\rightarrow \beta P_h$, where $\beta$ is a general constant, the equation
(\ref{7}) is left unchanged. We can advantage of this fact to choose the suitable normalization factor. For an expanding non-static universe
we fix $\beta=1/2$.
We obtain
\begin{eqnarray}
& & dU_h=\frac{c^4}{2G}dL_h+\frac{c^3}{2G}L_h^2dH+\frac{kc^4}{2G}\frac{L_h^3}{a^3}da,\label{8}\\
& & P_h=-\frac{kc^4}{4\pi G a^2}+\frac{3 c^3 H}{8\pi G L_h},\;T_{h}=\frac{\hbar c}{4\pi k_B L_h}.\label{9}
\end{eqnarray}
Thanks to this choice, the first term in (\ref{8}) is nothing else but the Misner-Sharp energy $E_{ms}=c^2 M_{ms}$, where $M_{ms}$ is the mass content within
the apparent horizon. 
Moreover, we expect that in the de Sitter case with $k=0$, since $L_h=const=c/H$, the effective pressure $P_h$ must be, apart from the sign, of the same value of
the one present in the energy momentum tensor 
$T_{\mu\nu}$, i.e. $P_h=|-\Lambda c^4/(8\pi G)|$
\footnote{Remember that the work term $pdV$ in the first law satisfies the usual rule: it is positive if the work is done by the system, while 
	is negative if the work is 'suffered' from the system.}. 

These facts justify this choice. 
It is also interesting to analyze the static limit for $H=0$ in (\ref{9}): this represents the static Einstein universe. In this case we expect
$P_h$ to be the effective pressure of this model i.e. $P_h=P_{eff}=p-\frac{\Lambda c^4}{8\pi G}$ where $p$ is the hydrostatic pressure present in
$T_{\mu\nu}$ and $P_{eff}$ the effective pressure acting in presence of a cosmological constant in the static universe. 
In this case $P_h$ is negative for $k=+1$ (the pressure must be negative to contrast the positive curvature) and obviously 
positive in the hyperbolic case. It is easy to see that by choosing the normalization factor $\beta=1/4$ we have
$P_h=\frac{-kc^4}{8\pi G a^2}, a=const$ and as a consequence by using $3k/(8\pi G a^2)={\rho}_{eff}=\rho+\Lambda c^2/(8\pi G)$, the other relevant
Friedmann equation $4\pi G/3(3P_{eff}+{\rho}_{eff}c^2)=0$ is satisfied. 

\section{The internal energy at the apparent horizon}

Once the quantities $P_h, U_h, E_{ms}, T_h$ have been identified, we can write the entropy formula (\ref{4}) in terms of these quantities. In fact, after
calculating $T_h S_h$ (with $\beta=1/2$), we have:
\begin{equation}
T_h S_h=\frac{E_{ms}}{2}+P_h V_h+\frac{k c^4 L_h^3}{16 G a^2}.
\label{10}
\end{equation}
Note that the expression (\ref{10}) generally does not assume the usual form $TS=U+PV$.

\subsection{The flat case}

First of all, we analyze in more detail the flat case $k=0$ obtaining $T_h S_h=E_{ms}/2+P_h V_h$, that in the static case $H=0$ reduces to the usual
black hole relation $T_h S_h=E_{ms}/2$.
With $\beta=1$ we obtain
$T_h S_h=E_{ms}+P_h V_h$ that is identical to the Euler's relation provided that $E_{ms}=U_h$. In the usual approach this is the key assumption,
but in our new approach $E_{ms}\neq U_h$. This happens because the gravitational contribution due to the expansion of the universe must be
included in $U_h$. As a consequence of our assumptions, by putting $k=0$ in (\ref{8}) we obtain, as calculated in \cite{8a}, 
$dU_h=0$, i.e. $U_h=const$. As noticed in
\cite{8a}, from dimensional arguments we have $U_h=0$. This result is in agreement with an old conjecture \cite{22} concerning the origin of the 
universe. In the flat case, the added term in (\ref{4}) proportional to $HV$ takes into account the gravitational degrees of freedom due to 
the expansion of the universe. The term $\sim dH$ in (\ref{8}) is the contribution to the internal energy due to the non-static
nature of our universe that in the flat case conspires with the energy of the matter to obtain a zero energy for the universe.
The internal energy $U_h$ is the energy necessary to create the system and thus it is not a surprise that in a 
spatially flat system this energy is zero, since our universe can be emerged from a minkowski spacetime. 

Concerning the enthalpy $\tilde{H}$ we have, by definition, ${\tilde{H}}_h=U_h+P_h V_h$. Since $U_h=0$, from (\ref{9}) we obtain
${\tilde{H}}_h=P_h V_h=\frac{c^4}{2G}L_h=E_{ms}$. As a consequence, in the flat case,
the enthalpy is nothing else but the Misner-Sharp energy (rather than the internal energy $U_h$). Furthermore, since 
$U_h=E_{ms}+E_{grav}=0$, we can compute the gravitational expansion energy $E_{grav}$ as $E_{grav}=-P_h V_h$. In the de Sitter
case, thanks to the reasonings at the end of the section 3, we have $P_h=-p=\Lambda c^4/(8\pi G)$ and 
$p V_h=-E_{ms}$, i.e. $p=-\rho c^2$, the correct result.

\subsection{General case}
In this subsection we analyze the most general case with $k\neq 0$, where $dU_h$ is not more identically zero. After writing the
(\ref{8}) as a time derivative we have:
\begin{equation}
U_{h,t}=\frac{c^4}{2G}L_{h,t}+\frac{c^3}{2G}L_h^2H_{,t}+\frac{kc^4}{2G}\frac{L_h^3}{a^3}a_{,t}.
\label{11}
\end{equation}
The internal energy (\ref{11}) is composed of three parts: the first is the usual Misner-Sharp contribution, the second can be associated to the 
expansion energy of the universe while the third term is the contribution due to the non-vanishing spatial curvature. The curvature
$k$ enters in all terms and conspires to give a non vanishing value for $U_h$. As pointed in \cite{8a}, this is a reasonable result since 
for an expanding universe with $k=-1$ we have the 'creation' of new negatively curved space, i.e. the creation of positive energy. Conversely,
to an expanding universe with $k=+1$ it is associated the creation of positively curved space, i.e. the creation of negative energy
(strong attractive regime). This interpretation is enforced by the fact that the Friedmann equations can be written in terms of the density
parameters ${\Omega}_m=\frac{8\pi G\rho}{3H^2}$, ${\Omega}_{\Lambda}=\frac{\Lambda}{3H^2}$, ${\Omega}_k=-\frac{kc^2}{a^2 H^2}$:
\begin{equation}
{\Omega}_m+{\Omega}_{\Lambda}+{\Omega}_k=1.
\label{12}
\end{equation}
Friedmann equations permit us to associate to the curvature term $-kc^2/a^2$ a binding energy described by the third term 
$\sim k$ in (\ref{11}). 

As stated above, in the flat case $U_h=0$. It is interesting to explore the existence of other spacetimes where the internal energy is conserved.
From (\ref{11}) we have
\begin{equation}
U_{h,t}=
\frac{c^5\left[H_{,t}\left(\sqrt{H^2+\frac{kc^2}{a^2}}-H\right)+2c^2ka_{,t}\right]}
{2Ga^3{\left(H^2+\frac{kc^2}{a^2}\right)}^{\frac{3}{2}}}.
\label{13}
\end{equation}
Note that the expression (\ref{13}) is not defined in the Milne case where the denominator is vanishing, implying $L_h=\infty$. Apart from this case, we
impose the equation $U_{h,t}=0$. From (\ref{13}) we deduce
\begin{equation}
H_{,t}\left(\sqrt{H^2+\frac{kc^2}{a^2}}-H\right)+2c^2ka_{,t}=0.
\label{14}
\end{equation}
To solve equation (\ref{14}) we use a simple argument. In a Friedmann context both $H$ and $a$ are time functions only. Hence in a monotonic 
(also decreasing with $H<0$) phase we have $H=H(t(a))$. Equation (\ref{14}) becomes, with the physical requirement $a>0$,
\begin{equation}
H_{,a}=-\frac{2c^2 k}{a^2\left[\sqrt{a^2 H^2+kc^2}-aH\right]}.
\label{15}
\end{equation}
By taking a general expression for $H(a)=G(a)$, both sides of (\ref{15}) are functions of $a$ only and the differential equation becomes an
algebraic constraint for $a$, i.e. $a=const$, but necessarily $H=0$. We thus have the static 
Einstein universe with $H=0, a=const$ with matter content such that $4\pi G(c^2\rho+3p)-\Lambda c^4=0$.
Thanks to these calculations, the total internal energy of this spacetime is zero.\\ 
By inspection of (\ref{15}),
the only time dependent solution is $H=Q/a, Q=const$. After putting this solution in (\ref{15}), it is easy to see that
no solution arises with $a>0$.
Hence, apart from the static Einstein solution, 
we found that the spatially flat solutions are the only Friedmann solutions with vanishing internal energy. In particular, the universe perhaps we live on 
average
has zero total energy. When a non-vanishing curvature is present, this curvature gives a contribution to the energy that is negative for $k>0$
(strong attractive regime, negative binding energy) and positive for $k<0$ (positive binding energy, repulsive regime). 

Consider now Friedmann universes filled with an ordinary perfect fluid together with a positive cosmological constant $\Lambda$. As well known, these solutions admit an asymptotic behaviour given by the de Sitter metric.
From (\ref{13}) we can calculate 
the leading term at late times for $t\rightarrow\infty$:
\begin{equation}
U_h=-\frac{kc^7}{2Ga^2 H_{\Lambda}^3}+o(1/a^2),\;\;\;H_{\Lambda}=c\sqrt{\frac{\Lambda}{3}},
\label{16}
\end{equation}  
where the integration constant has been set to zero since the energy of the de Sitter universe is zero. Note that by defining the density 
${\rho}_k$ by $c^2{\rho}_k V_h=U_k$, from (\ref{16}) we have for $t\rightarrow\infty$
\begin{equation}
{\rho}_k\rightarrow -\frac{3kc^4}{8\pi G a^2}
\label{il}
\end{equation}
that is exactly the density related to ${\Omega}_k$ in (\ref{12}).\\
In practice, from (\ref{16}) we see that this asymptotic behaviour looks like the binding term $-kc^2/a^2$ present in the Friedmann equation
(\ref{12}). We thus obtain that Friedmann universes equipped with a positive cosmological constant all have an internal energy 
asymptotically (from the below for $k=-1$ and from the above for $k=+1$) approaching a zero value.

\section{The free Helmholtz energy}

The Helmholtz energy $F=U-TS$ is a thermodynamic function that, in ordinary systems, is 
stationary for all systems in thermal equilibrium with a given 'reservoir'. The stationary points of $F$ (at constant volume) represent
isothermal thermodynamics equilibrium states. As well known \cite{hen,w1}, the Helmholtz energy 
can be
defined also in nonequilibrium states. 
In the usual picture where $U=E_{ms}$, $S\sim A/4$ and $T$ is given by (\ref{6}), we found $F=0$. If this were the case, any
Friedmann universe should be in thermal equilibrium at the apparent horizon, 
but we expect that $F_{,t}=0$ for $T=const$ at $V=const$: in the cosmological context 
this implies $L_h=const$, i.e. a Sitter universe. In the following we investigate the expression for $F_h$ arising from our formulas.
We obtain:
\begin{equation}
F_h=U_h-\frac{c^4 L_h}{4G}-\frac{c^3L_h^2 H}{2G}+\frac{k c^4 L_h^3}{4G a^2},
\label{17}
\end{equation}
with $U_h$ given by (\ref{11}) after a time integration. We study the monotony of $F_h$. For 
$F_{h,t}$ we obtain:
\begin{equation}
F_{h,t}=\frac{c^3}{G}L_{h,t}
\left[\frac{c}{4}-HL_h+\frac{3ck L_h^2}{4a^2}\right].
\label{18}
\end{equation} 
From (\ref{18}) if follows that $F_{h,t}=0$ for $L_h=constant$, i.e. for a de Sitter spacetime or for spacetimes equipped with a positive 
cosmological constant asymptotically approaching a de Sitter phase. First of all, for an ordinary expanding spacetime 
$L_{h,t}>0$, while for a de Sitter spacetime $L_{h,t}=0$. Only spacetimes equipped with a 'phantom' equation of state
($p=wc^2\rho, w<-1$) have $L_{h,t}<0$ ($L_{h,t}\sim (1+w)/\sqrt{\rho}$). 
In the following, we assume universes filled with a non-phantom fluid.
Moreover, note that both for $k=0$ and $k=-1$ certainly the expression between square brackets is always strictly negative. 
We thus obtain that for $k=0,-1$ $F_h$ is always a non-increasing time function and is stationary only when a de Sitter phase is reached.\\
Concerning the case with $k=+1$, we note that the expression in square brackets is vanishing only for $Ha=2c/\sqrt{5}$. Moreover,
$F_{h}$ is decreasing for $Ha>2c/\sqrt{5}$ and increasing for $Ha<2c/\sqrt{5}$. However, we expect that our calculations
for $k=+1$ (and also the entropy bound, see \cite{21}) to obtain $S_h$ are certainly valid for $L_h<<a$ and in this range certainly we have
$F_{h,t}<0$. Also for $k=+1$ but with $\Lambda>0$ the term in brackets is expected to be negative. 
As a result, also for the closed case as far as our assumptions are certainly valid, the free energy is monotonically decreasing and
stationary only when $L_h=const$. These results are in agreement with the physical meaning of the free energy.

\section{Holographic and thermodynamic temperature of the universe}

In the sections above we have studied the entropy, the internal energy and the free energy of the universe by invoking the holographic principle.
In this context, the 'holographic' temperature $T_h$ at the apparent horizon arises. 
However, this holographic temperature is not the one $T_u$ of the universe present
inside the apparent horizon. It is clearly not so obvious to define an average temperature $T_u$ for the whole universe. Since 
the highly homogeneous and isotropic matter-energy content of the universe is provided by the CMBR, we may identify with good approximation 
$T_u$ with the temperature $T$ of the CMBR. At the present epoch $T_0\simeq 2.72\;K$ and it scales in the following way:
$T_u(t)=T_0/a(t)$. In the following, we investigate the physical relation between $T_h$ and $T_u$. To this purpose, the fundamental result is the 
one obtained in section 5, equation (\ref{17}): the free Helmholtz energy is stationary only when a de Sitter phase is reached,
i.e. $L_h=const$. Since the free energy is stationary only for systems in thermal equilibrium, during a de Sitter phase
for the holographic $T_h$ and thermodynamic temperature $T_u$ we must have 
\begin{equation}
T_h=\frac{\hbar c}{4\pi k_B L_{\Lambda}}=T_u,\;\;L_{\Lambda}=\sqrt{\frac{3}{\Lambda}},
\label{24}
\end{equation}
for any positive cosmological constant $\Lambda$. This means that to a cosmological constant must be associated a non-vanishing temperature.
This temperature is vanishing in the limit $L_h\rightarrow\infty$ and its non zero value is a finite size effect induced by the
finite thermodynamic radius $L_h$ of our universe. 
By continuing on this line of reasoning, according to physical intuition and to the physics of an ordynary gas, it is natural to suppose
that when $T_u>T_h$ the apparent horizon of our universe is expanding 
(i.e. $p=w c^2\rho, w>-1$), while when $T_u<T_h$ we have a phantom (cold) universe ($w<-1$) with a decreasing
apparent horizon, i.e.  
when $T_u>T_h$ we expect an expanding apparent horizon with $L_{h,t}>0$. Conversely, when
$T_u<T_h$ we expect a contracting apparent horizon with $L_{h,t}<0$. Soon after the big bang near the Planck era we expect
$T_u-T_h>0$, i.e an expanding universe with deceleration parameter $q>-1$. 
After the Planck era, the effective temperature of the universe $T_u$ quickly decreases
until $T_u-T_h\simeq 0$ and a pure de Sitter inflationary era begins.  At late times, an asymptotically pure de Sitter phase arises for
$T_u\rightarrow T_h$. This picture can be verified by numerical estimations.\\    
At the present epoch, we have $T_{0u}\simeq T_{0CMBR}\simeq 2.72\;K$, while $T_{0h}\simeq 1.76\times 10^{-29}\;K$ and thus
$T_u>T_h$. Asymptotically we have $T_h=T_u\simeq 1.2\times 10^{-29}\;K=T_{\Lambda}\; $ and the so called thermal death of the universe does not 
arise in our view. 

At the Planck era ($t\simeq t_P$) it is natural to suppose $T_u=T_P=\sqrt{\hbar c^5/(G k_B^2)}$.\\ 
Below the Planck time the notion of classical manifold breaks down. 
We may for example suppose a non-commutative structure of the spacetime (quantum spacetime \cite{dop}). In this context, it is expected 
(see \cite{23,24}) that a maximum finite value for $H$ is attained of the order of $H_{max}\sim 1/(\sqrt{3}t_P)$ and we expect that the inequality
$T_u>T_h$ still holds. At the Planck era and before inflation the matter content of the universe can be represented by a radiation field at the temperature
$T_u$ scaling as $T_u=T_i/a(t)$ in a spatially flat Friedmann universe with $a(t)=\sqrt{\frac{t}{t_i}}$. By taking for a practical estimation
for $t_i$ and $T_i$ the actual time and the actual temperature of the CMBR 
($T_i$ and $t_i$ must be calculated at the same cosmic time) with $H=1/(2t)$ we have that the inequality
\begin{equation}
\frac{\hbar H}{4\pi k_B}\leq\frac{T_0}{a(t)}
\label{25}
\end{equation}   
is satisfied for $t\geq t^*\simeq 0.2 t_p$. Thanks to the non commutative effects below the Planck time, we have that the maximum value for $H_{max}$
is reached for $t\simeq 0.86 t_P$,
i.e. we can reasonable assume the validity of the assumption $T_u-T_h\geq 0$ $\forall t>0$.\\
At the beginning of the primordial inflation with $H=H_I$, we must have 
$T_u=T_{hI}=\frac{\hbar H_I}{4\pi k_B}$. After an unknown reheating mechanism the inflation ends and $T_u >T_h$ up to 
late times where $T_u\rightarrow T_h$. 

Concerning $T_u$, one may think that the main contribution to $T_u$ is due to the CMBR radiation at the temperature $T=T_0/a(t)$ with 
density ${\rho}_{r}$, with a correction given by the temperature $T_{\Lambda}$ of the cosmological constant $\Lambda$. Moreover, we 
label with $T_i$ all the remaining contributions to $T_u$ with density
${\rho}_i$ (for example the contributions from megnetic fields, neutrinos, axions...).  
At first look and for practical purposes, we can consider the universe as a non decoupled
mixture between the radiation ${\rho}_r$ and the dark components of the universe
${\rho}_{\Lambda}=\Lambda c^2/(8\pi G), {\rho}_i$. We thus have within the apparent radius $L_h$
\begin{equation}
T_u=\frac{{\rho}_{\Lambda}T_{\Lambda}+{\rho}_r T_r+{\rho}_i T_i}{{\rho}_{\Lambda}+{\rho}_r+{\rho}_i}.
\label{29}
\end{equation}
For universes with a positive cosmological constant, we expect that asymptotically 
$\{{\rho}_r,{{\rho}_i}\}\rightarrow 0$ and as a result 
$T_u\rightarrow T_h\rightarrow T_{\Lambda}=\hbar\sqrt{\Lambda}/(4\pi k_B\sqrt{3})$.

An interesting consequence of our formulas is that to a cosmological constant in an 
expanding universe can be attributed a non-vanishing temperature $T_{\Lambda}$. 
This happens because for an expanding universe filled with
non-phantom matter-energy, the free energy is not increasing and is stationary only when a de Sitter phase is reached.
Since the free energy is stationary only for systems in thermal equilibrium, a de Sitter non-static universe
can be depicted as a thermodynamical system
such that the universe at $L_h$ is in thermal equilibrium at the temperature $T_u$ 
with the 'reservoir' at the holographic temperature $T_h$ with $T_u=T_h$. 

Summarizing, suppose that we have an expanding Friedmann universe filled with a certain set $\{s\}$ of matter-energy fluids 
with equation of state $p_s={w}_s{\rho}_s c^2, w_s\geq -1$ and 
with temperature $T_s$.
Also suppose that the different kinds of matter present in the universe are not decoupled and as a consequence
a mean temperature $T_u$ for the universe can be defined.
Hence, for the whole history of such a universe we have $T_u\geq T_h$, where
\begin{equation}
T_u=\frac{\sum_s {\rho}_s T_s}{\sum_s {\rho}_s}\geq \frac{\hbar c}{4\pi k_B L_h},
\label{31}
\end{equation}
We can manipulate the formula (\ref{31}) in the following way. For the homogeneity and isotropy of Friedmann universes, all the matter-energy
density components of the universes are time functions only. To this purpose, we denote with $E_{ms}^s$ the general Misner-Sharp mass of any
given component of the universe at $L_h$.\\ 
After using the Friedmann equation $H^2+kc^2/a^2=8\pi G/3\sum_s{\rho}_s$, the (\ref{31}) becomes
\begin{equation}
\sum_s T_s E_{ms}^s\geq \frac{\hbar c^5}{8\pi k_B G}.
\label{32}
\end{equation}
Equation (\ref{32}) looks like the usual time-energy inequality of ordinary quantum mechanics but with the temperatures instead of the time,
while inequality (\ref{31}) as a kind of temperature-length uncertainty relation.
The physical meaning of (\ref{32}) (an also of (\ref{31})) is very simple. 
Suppose for simplicity a universe dominated by a single matter energy component '$k$'.
If the Misner-Sharp mass of this component were very low, its temperature should be huge. In effect, near the big bang we expect
$E_{ms}^k\rightarrow 0$ and $T_k\rightarrow\infty$. It should be noticed that at the big bang $S_u=0$ and $S_{u,t}\geq 0$ for non-phantom
universes and as a consequence the second law of thermodynamic is satisfied for our
universe at the apparentt horizon, as well known, only  for non-phantom universes. 
In this regard, the inequalities (\ref{31}) and (\ref{32}) are in agreement with the second law of thermodynamics.

Note that, as stated in this section, inequality (\ref{31}) or (\ref{32})
are satisfied above the Planck time, but below the Planck
time $t_P$, inequality (\ref{32}) can be violated. In fact, for a power low cosmology with radiation-filled matter, we have 
$T_r\sim 1/a(t)$ and $a(t)\sim \sqrt{t}$. Hence $T_r E_{ms}^r\sim \sqrt{t}$ and (\ref{32}) is violated below $t\simeq t_P$.
However, we expect that the spacetime cannot more be depicted as a classical manifold but rather that the non-commutative quantum 
structure of the spacetime emerges (see \cite{dop,23,24,24b,d1,d2,pia,dop2}). In this frame, it is expected that a minimum volume does appear
(see in particular \cite{d2,23,24}) of the order of the Planck volume and as a result the inequalities is not violated.
This implies that a quantum spacetime conspires belove the Planck time to save the second law of thermodynamics for a non-phantom filled universe.
This implies that for spatimes filled with a non phantom matter content, the second law is always satisfied.

\section{Discussion and final remarks}

In this paper we have investigated some interesting thermodynamical consequences of our new recently proposed formula for the Bekenstein-Hawking
entropy in expanding universes. This formula, according to the holographic principle, has been applied to the apparent horizon of our visible universe.
We showed that only flat Friedmann universes have zero internal energy. The introduction of a non-vanishing
spatial curvature $k$ inhibits the conservation of the total energy of the universe. This can be explained by the fact that, thanks to
the Friedmann equations, to a non-zero spatial curvature can be associated a 
binding energy. For an expanding universe with $k=-1$ new space is created with positive energy, while for $k=+1$ with negative energy
(attractive gravity).

Another interesting and useful result of this paper is the one of section 5 regarding the free Helmholtz energy $F_h$: 
the free energy of the universe is stationary only when a de Sitter phase is reached.  
The first implication is that to a cosmological constant 
of an expanding universe for comoving observers can be associated 
a non-vanishing temperature that in turn is the mean temperature of the universe during a pure de Sitter phase. 
Since the free energy is stationary
only when a de Sitter phase is reached, we may think our universe as evolving between two equilibrium phases.
Hence, the actual small cosmological constant allows the universe to evolve up to an asymptotically stable equilibrium configuration.
Also note that in this new view, the primordial inflation happens because the temperature $T_u$ of the universe dropped
more quickly with respect to $T_h$, and as a consequence a pure de Sitter phase begins when $T_u=T_h$. This interesting point of view has been 
partially addressed at the appendix of \cite{14}.
Our view can have some similarity with the construction made in \cite{25}, where the Einstein's equations have been written in such a way that the non-static
nature of a given spacetime with respect to a suitable local reference frame is due to the difference between the degrees of freedom
$N_{sur}$ of the boundary of a given bulk region and the ones of a bulk $N_{bulk}$. In this context, the degrees of freedom are computed in terms
of the average Davies-Unruh temperature $T_{avg}$ at the boundary of a bulk region. Although some interesting analogy can exist 
(instead of $N_{sur}-N_{bulk}$ we use $T_u-T_h$ to describe the thermal history of our universe), we refer to the temperature of the
bulk $T_u$ as the temperature of the matter-energy content inside $L_h$ at the present time, (practically the temperature of the CMBR), 
while $N_{bulk}$ in \cite{25} is defined in terms of $T_{avg}$. In our view the difference $T_u-T_h$ is relevant for the dynamics of the proper
areal radius $L_h$. A de Sitter universe with $T_u=T_h$ has an apparent horizon with stationary proper area, but the universe is still
expanding for a non local reference frame (comoving observers). Moreover, in \cite{25} it is assumed the usual equipartion law, but this is 
strictly valid for systems whose Hamiltonian is quadratic in the conjugate variables and in the classical range where the fluid is described in the
canonical ensemble by the Boltzmann distribution. 

Further interesting physical consequences are due to the inequalities (\ref{31}) and (\ref{32}). The inequality (\ref{31}) in particular can be put
in a form similar to a quantum uncertainty relation for the temperature $T_u$ and the proper radius $L_h$ of our universe as
\begin{equation}
T_u L_h\geq \frac{\hbar c}{4\pi k_B}.
\label{33}
\end{equation} 
Since at first look the quantities at the left member of (\ref{33}) have nothing to do with the fundamental constant 
$\hbar$, the presence of $\hbar$ in (\ref{33}) is rather intriguing. For our universe, homogeneous on scales $>70-100$ Mpc, we can consider
$T_u$ as an average quantity $<T_u>$ calculated with respect to different inhomogeneous patches.

An important consequence of (\ref{33}) (or (\ref{32})) is that a given 
Friedmann universe filled with non-phantom mass-energy at any fixed $L_h$ cannot have an arbitrary small temperature.
Only for expanding spacetimes without a 
cosmological constant
we have that when $t\rightarrow\infty$, $L_{h}\rightarrow\infty$ and then the minimum for $T_u$ can be zero, but only asymptotically.
It should also be noticed that this result has not been obtained by the use only of Friedmann equations, but with the help of the
behaviour and meaning of the free energy, equation (\ref{18}). A standard result, obtained after writing the Friedmann equations 
as a first law with $S_h\sim A_h/4$ but without 
the extra imput of the free energy can be found in \cite{26}. In this paper it is shown that, by supposing that $T_h$ is proportional to
$T_u$ and the generilezed second law holds,
$T_h-T_u\leq Q$, where $Q\sim (\rho c^2+p)$.\\ 
For $c^2\rho\geq -p$ (non-phantom matter), we have 
\begin{equation}
T_u-T_h\geq -Q,\,\; Q\geq 0.
\label{34}
\end{equation}
From (\ref{34}) we argue that for a non-phantom universe no 
positive minimum allowed value
for $T_u$ emerges. In our calculations, the inequality (\ref{33}) is only a consequence of the physical meaning of the free energy. 
Hence, thermodynamics applied at the whole visible universe can hint us the nature of
the expansion of the visible universe. 

In our view the (proper area of) apparent horizon is time-increasing if and only if $T_u>T_h$, decreasing for $T_u<T_h$ and stationary for $T_u=T_h$. 

A dark energy can thus be interpreted as
a perfect fluid satisfying the thermodynamic relation $T_h L_h=const$ or
$T_h^3 V_h=const$. According to the stationary isoentropic nature of the apparent horizon of a de Sitter expanding universe,
the equality $T_h^3 V_h=const$ is nothing else but the equation of an adiabatic isoentropic transformation with
$S_h\sim V_h T_h^3$.

It is important to note that inequality (\ref{33}) is independent on the gravitational constant $G$. This fact is in line with the holographic principle
where the entropy bound does not depend on $G$ an thus can be considered as universal. This suggests the possibility that also the 
inequality (\ref{33}) is universal and can be applied to any physical probe. We can thus formulate the following conjecture:

{\it Any given matter-energy fluid of volume $V$ and satisfying
a non-phantom equation of state, must satisfy inequality (\ref{33})}.\\ 
As mentioned above, the inequality (\ref{33}) can be written in terms of the 
volume:
\begin{equation}
T_{u}^3 V \geq \frac{1}{48{\pi}^2}{\left(\frac{c\hbar}{k_B}\right)}^3.
\label{al}
\end{equation}
When non commutative effects are taken into account (see \cite{d2,23,24}), a minimum volume does appear and as a result a maximum temperature 
$T_{umax}$
emerges of the order of the Planck temperature $T_P$, i.e. $T_{umax}\sim T_P/(4{\pi})$ (big bang is ruled out).\\
Conversely, for a phantom gas the minimum attainable value 
(but not the maximum) for $T$ disappears and thus the (\ref{33}) or (\ref{al}) can be violated. In particular, for
a phantom matter probe with temperature $T_{ph}$ and length $L$, we must have 
\begin{equation}
T_{ph} L < \frac{\hbar c}{4\pi k_B},
\label{35}
\end{equation} 
or the reverse of (\ref{al}).
Note that our proposal for a non vanishing value $T_h$ for the temperature of the dark energy (see for example
\cite{sar} and references therein) allows a non negative value for the temperature of a phantom fluid given by
(\ref{35}) thus solving the shortcomings of a negative temperature. 
Nevertheless, since $L\geq 0$, we can in principle have a phantom probe with an absolute
temperature $T_{ph}<0$ (i.e. $T_{ph}<-273.15$ C) (see \cite{27}). 

Summarizing, our conjecture implies that for a given gas in a box of typical size $L$ there exists a critical temperature $T_c=\frac{\hbar c}{4\pi k_B L}$.
Above this temperature, the gas has not a phantom-like behaviour, below $T_c$ the system is in practice a phantom-gas and exactly at
$T=T_c$ the system looks like a cosmological constant. This critical temperature depends on the size of the probe:
a given gas at the temperature $T$ can be in a phantom or non phantom phase depending on its size according to inequality
(\ref{33}) and (\ref{35}). For a numerical example, for $L\sim 1$ meter, i.e. a box of length $1$ meter, we have $T_c\simeq 0.00025 K$, 
for $L\sim 1$ centimeter, $T_c\simeq 0.025 K$, while for the whole universe (apparent horizon)
$T_c\simeq 10^{-29} K$.

Obviously, independently on the correctness of the conjecture above for non cosmological systems,
all these reasonings can be of interest for a better understanding on the nature of the dark energy.

As a final consideration,
note that in the static case, the black hole temperature satisfies 
$T_{bh}M_{bh}=\hbar c^3/(8\pi G k_B)$, i.e. in the case of a thermal equilibrium the inequality (\ref{al}) becomes an equality identical to the one of a
static black hole. This means that a de Sitter expanding universe at $L_h$ has the same temperature-mass relation of a static black hole
with the Misner-Sharp mass instead of the ADM mass. 
This is not surprising since in an expanding universe, thanks to \cite{10,11}, a black hole can be identified by its apparent horizon rather than the
teleological event horizon. In the static case
the time coordinate $t$ (as measured by the observer at spatial infinity)
becomes spacelike and the spatial coordiante $r$ timelike inside the event horizon 
as happens for the de Sitter 
expanding universe. This hints the possibility that also the interior of a static black hole can be filled with a cosmological-like
dark energy at the temperature $T_{bh}$ of its horizon. 
This hypothesis can be matter of future investigations.

\section*{Appendix a}

McVittie \cite{mac} discovered over 80 years ago a class of solutions in the attempt to describe the metric of a spherical object embedded in an
expanding Friedmann universe. The interpretation of this solution as describing a black hole in an expanding universe is currently
debated (see for example \cite{mac1} and references therein). Only recently \cite{mac1,mac2} it
has been shown that the spacetime contains a surface
(a null apparent horizon) that in the far future limit $t\rightarrow\infty$ and in presence of a constant positive limit for the Hubble flow
$H$ looks like a regular event horizon. The line element for the McVittie solution can be written in isotropic coordinates as follows \cite{mac2}:
\begin{equation}
ds^2=-\frac{{\left(1-\frac{Gm}{2c^2a(t)r}\right)}^2}{{\left(1+\frac{Gm}{2c^2a(t)r}\right)}^2}c^2 dt^2+
a(t)^2{\left(1+\frac{Gm}{2c^2 a(t)r}\right)}^4(dr^2+r^2d{\Omega}^2),
\label{a1}
\end{equation}
where $m$ is a mass parameter and $H(t)=a_{,t}(t)/a(t)$. For the solution (\ref{a1}) we have $K_{a}^{a}=3H$ that is obviously constant on  
comoving foliations
and the conditions to apply formula (\ref{3}) are fulfilled. Moreover, for the density $\rho(t)$ we have the usual Friedmann equation
$H^2=8\pi G\rho/3$. We can investigate the relation between the mass parameter $m$ of (\ref{a1}) with the mass excess parameter
$\delta M$ present in (\ref{3}). To start with, note that the McVittie solution reduces to a Friedmann flat universe for $m=0$ and as a consequence 
we limit to the case $k=0$ in (\ref{3}). By inverting the inequality (\ref{3}), we have a sufficient condition for the formation of trapped surfaces.
After denoting with $M_d$ 
\begin{equation} 
\frac{G}{c^2} M_d = \frac{L}{2}+\frac{AH}{4\pi c},
\label{a2}
\end{equation} 
this condition becomes $\delta M\geq M_d$. Remember that $\delta M$ denotes the proper mass excess necessary to form a trapped
surface. At this stage of the tractation, the Bekenstein bound comes in action, i.e. $S\leq S_{max}=\frac{2\pi k_B L E}{\hbar c}$.
In the usual tractation, $E$ is the energy related to the mass concentration leading to a black hole. In the static context, the term with 
$H$ is vanishing in (\ref{a2}) and inequality $\delta M\geq M_d$ is saturated with the ADM mass $m_{adm}$ with $m_{adm}=c^2 L/(2G)$.
In a cosmological context, where we have not an asymptotically flat 
reference metric, the situation is rather more complicated and it is not clear what kind of energy $E$ is suitable for 
the Bekenstein bound.
The theorems quoted in \cite{15,16,17,18} suggest that in a cosmological context the correct energy to put in the Bekenstein bound
is the ('dynamical one') $E=M_d c^2$.

The McVittie solution (\ref{a1}) has apparent horizons at the proper areal radius 
$L=r a(t){\left(1+\frac{Gm}{2c^2 a(t)r}\right)}^2$ given by the roots of 
the equation involving ingoing geodesics:
\begin{equation}
\frac{H^2 L^3}{c^2}-L+\frac{2Gm}{c^2}=0.
\label{a3}
\end{equation}
Equation (\ref{3}) has roots depending on the discriminant $D=-1/(27 H^6)+m^2/(4H^4)$. Only for $D\leq 0$ we have positive roots. We denote
with $L_A$ the smaller positive root. The surface $L=L_A$ at a fixed time is a spacelike apparent horizon that for $t\rightarrow\infty$
and for $H\rightarrow H_0=const>0$ in this limit becomes the future event horizon and a black hole emerges.

The mass parameter $m$ in (\ref{a1}) is not the active gravitational mass. The active gravitational mass $M_A$ calculated at 
$L_A$ is given by $M_A=m+\rho V_A$, where $V_A=4\pi L_A^3/3$ and hence by the equation $H^2=8\pi G\rho/3$ we have
\begin{equation}
M_A=m+\frac{H^2 L_A^3}{2G},
\label{a4}
\end{equation}
and thanks to the (\ref{a3}) we get $M_A=\frac{c^2 L_A}{2G}$, i.e. $M_A$ is nothing else but the Misner-Sharp mass $M_{ms}=M_A$.
As a consequence for the McVittie solution (\ref{a1}) we have
\begin{equation}
M_d=M_{ms}+\frac{c H}{G}L_A^2.
\label{a5}
\end{equation}
According to (\ref{3}), to form a black hole in an expanding universe it requires a proper mass concentration
$M_d$ greater than the Misner-Sharp mass $M_{ms}$\footnote{Note that the Misner-Sharp mass is expected to include the negative 
	gravitational energy inside $L_A$.} 
by a term depicting the dynamical nature of a Friedmann cosmological
background. In practice, for spherical objects with areal radius $L_A$
embedded in a Friedmann flat universe, the Bekenstein entropy bound can be 
generilized to encompass the dynamical degrees of freedom due to the expanding universe in the following way:
\begin{equation}
S\leq S_{max}=\frac{2\pi k_B L_A}{\hbar c}\left[E_{ms}+\frac{c^3 H}{G}L_A^2\right], 
\label{a6}
\end{equation} 
with $L_A$ the areal radius of the spherical region and $E_{ms}=M_{ms}c^2$ the Misner-Sharp energy of a black hole at its apparent horizon
$L_A$.

As a final consideration, we could apply the technology of paper \cite{8a} to calculate the first law of thermodynamics for
(\ref{a1}) at $L_A$. We obtain in particular an expression for the internal energy $U_A$ of (\ref{a1}) at $L=L_A$ given by
\begin{equation}
U_{A,t}=\frac{c^4}{2G}L_{A,t}+H_{,t}\left[\frac{c^3}{2G}L_A^2+
 \frac{4\pi}{c}L_A^4 T_A^{(2)}\right],
\label{a7}
\end{equation} 
where $T_A^{(2)}$ is the trace of the energy-stress tensor on the normal space $\{0,1\}$ of (\ref{a1}) calculated at $L_A$.
It can be interesting to note that in presence of a positive cosmological constant where the spacelike apparent horizon 
$L_A$ becomes for $t\rightarrow\infty$ the future event horizon, the internal energy is conserved in such a limit, i.e. the internal energy 
of the teleological McVittie black hole is constant in time. Since for $m=0$
\footnote{As shown in \cite{8a}, the internal energy of a Friedmann flat universe equipped with a positive 
	cosmological constant at the apparent horizon is zero.}
the (\ref{a1}) reduces to the Friedmann flat metric and that in presence of a positive cosmological constant $\Lambda$ 
$L_A(t\rightarrow\infty)\sim 1/\sqrt{\Lambda}$, 
we conclude that the total internal energy $U_A$ for the 
McVittie black hole is $U_A=\gamma mc^2,\;\gamma\in  R^{+}$ and not $M_{ms}c^2$. By choosing $\gamma=1$, we can identify the mass
parameter $m$ as the total internal energy of the teleological McVittie black hole.

\section*{Appendix b}

In this appendix we study the partition function in the flat case, where a simple expression for $F_h$ is available. 
In this case we have 
\begin{equation}
F_h=-T_h S_h=-\frac{3c^4}{4G}L_h=-\frac{3c^5}{4GH}.
\label{19}
\end{equation}
To start with, we must formally introduce the partition function $Z$. Suppose that the spacetime is equipped with a set $\{s\}$ of countable
microstates. 

The partition function can thus be obtained in the standard way by means of the expression 
$S=-k_B\sum_s p_s\ln p_s$ together with the constraints $\sum_s p_s=1, <U>=\sum_s p_s U_s$, where $p_s$ is the probability  and $U_s$ the energy
of the microstate '$s$'. Thanks to usual relations and the technique of Langrangian multipliers we get
\begin{equation} 
Z=e^{-\frac{F}{k_B T}}=\sum_s e^{-\frac{U_s}{k_B T}},\;\;p_s=\frac{e^{-\frac{U_s}{k_B T}}}{Z},\;\;<U>=\frac{\sum_s e^{-\frac{U_s}{k_B T}} U_s}{Z}.
\label{20}
\end{equation}
In the flat case we found $<U_h>=0$. From (\ref{20}) we must have
\begin{equation}
<U_h>=\frac{\sum_s e^{-\frac{U_{hs}}{k_B T_h}} U_{hs}}{Z}=0.
\label{21}
\end{equation}
The sum in (\ref{21}) can be made zero by supposing that for any microstate '$i$' with energy $U_{hi}>0$, there exists a microstate
'$j$' with energy $U_{hj}<0$ such that $z_i U_{hi}=-z_j U_{hj}$ where $z_k=e^{-U_{hk}/k_BT_h}$, but we have no arguments supporting this 
condition. Hence, the most simple and conservative hypothesis is to set $\forall i\in \{s\}, U_{hi}=0$. As a consequence, thanks to
(\ref{19}) and (\ref{20}) we have 
\begin{equation}
S_h=k_B \ln Z,\;\;Z=\sum_s\;microstates. 
\label{22}
\end{equation}  
We can measure the proper radius $L_h=c/H$ of the apparent horizon in terms of the fundamental 
Planck length $L_P$, i.e. $L_h/L_P=N_P$. We obtain $S_h=k_B 3\pi N_P^2$ and
\begin{equation}
Z=\sum_s\;microstates=e^{3\pi N_P^2}.
\label{23}
\end{equation} 
According to the formulas (\ref{22}) and (\ref{23}), the Gibbs factor is exactly 1 and as a consequence the expression for 
$S_h$ reduces to the one of the microcanonical ensemble.

\end{document}